\documentclass[aip,apl,preprint]{revtex4-1}
\usepackage{graphicx}
\usepackage{dcolumn}
\usepackage{bm}

\begin{document}
\title{Electromechanical wavelength tuning of double-membrane photonic crystal cavities} 
\author{L. Midolo}
\email[]{l.midolo@tue.nl}
\author{P. J. van Veldhoven}
\author{M. A. Dundar}
\author{R. N\"{o}tzel}
\author{A. Fiore}
\affiliation{COBRA Research Institute\\Eindhoven University of Technology, P.O. Box 513, NL-5600MB Eindhoven, The Netherlands}

\date{\today}

\begin{abstract}
We present a method for tuning the resonant wavelength of photonic crystal cavities (PCCs) around 1.55 $\mu$m. Large tuning of the PCC  mode is enabled by electromechanically controlling the separation between two parallel InGaAsP membranes. A fabrication method to avoid sticking between the membranes is discussed. Reversible red/blue shifting of the symmetric/anti-symmetric modes has been observed, which provides clear evidence of the electromechanical tuning, and a maximum shift of 10 nm with $<6$ V applied bias has been obtained.
\end{abstract}

\pacs{42.60.Da,81.07.Oj, 85.35.Be, 81.05.Ea}
\maketitle 

Two-dimensional photonic crystal cavities (PCCs), fabricated on III/V materials, have been widely used as key components for the realization of many nanophotonic devices such as low-threshold lasers,\cite{15} filters and switches operating at telecommunication wavelengths. Moreover, when coupled to quantum emitters like quantum dots (QDs), PCCs are employed in the investigation of cavity quantum electro dynamics (CQED) phenomena,\cite{25} thanks to their large quality factors Q and small mode volumes. 
Unfortunately, small imperfections due to the fabrication process have a large impact on the optical properties of the cavities, making a deterministic production of devices impossible. Therefore a method to control the resonant modes (tuning) of PCCs in real time is required. 
Previous works already demonstrated several techniques to gain control over a photonic crystal device. These include thermo-optic tuning,\cite{16} carrier injection,\cite{7} infiltration with water\cite{17} and with liquid crystals,\cite{10} or approaching dielectric tips to the surface of the crystal.\cite{9,22} However most of these methods have their own drawbacks such as large degradation of Q factors, small tuning ranges, slow response or incompatibility with on-chip integration. Recently, tunable crystals have been realized on nano-electro-mechanical structures (NEMS),\cite{8,19} by controlling the lateral separation of two parallel nanobeams. In this work we report a tuning scheme based on coupled 2D PCCs fabricated on two parallel InGaAsP slabs (double membranes) whose distance in the vertical direction is controlled using electrostatic forces. As compared to the nanobeam approach, the use of out-of-plane actuation opens the opportunity to integrate the tuning control with an active region, by separating them in the vertical structure. The double-membrane photonic crystal structure has been already theoretically proposed\cite{1} and experimentally demonstrated\cite{12} but electromechanical actuation has not yet been reported.

\begin{figure}
\includegraphics{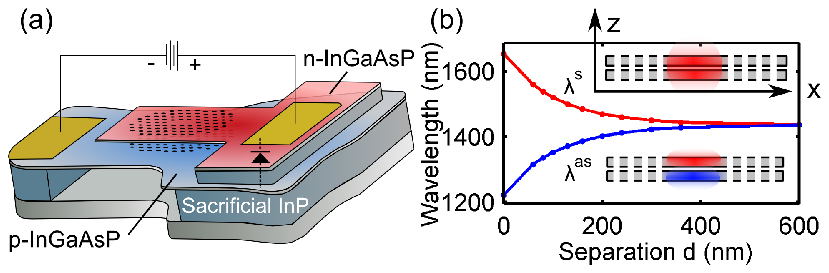}%
\caption{\label{fig:theory}(a) Schematic representation of the electrostatically tunable PCC on double membrane. (b) FEM simulations of the PCC mode tuning as a function of the inter-membrane separation $d$.}%
\end{figure}

To displace the membranes in a controllable and reversible way, electrostatic forces can be used by applying an electric field between them. To generate a uniform field across the air spacing $d$, two doped layers may be employed as done in electromechanically tunable vertical-cavity lasers.\cite{2} By forming a p-i-n junction as shown in Figure \ref{fig:theory}a and operating it under a reverse bias V, an electrostatic pressure $\varepsilon_0\frac{V^2}{2d^2}$ (assuming an ideal capacitor, i.e. neglecting the voltage drop in the depletion regions) can be exerted on both slabs. Such a pressure, in the range of 0.1--1 kPa, is sufficient to move a free-standing structure such as a cantilever (Fig.\ref{fig:theory}a) or a doubly clamped beam. 

The three-dimensional mode profile of PCCs on single slabs can be separated into an in-plane component, which depends on the geometry of the photonic crystal, and an out of plane component determined by the wave-guiding conditions. The effective refractive index $n_{\text{eff}}$ of the guided mode relates the wavelength of the cavity emission in free-space ($\lambda_0$) to the one in the slab by $\lambda_0=\lambda n_{\text{eff}}$.
When two identical membranes are brought at close distance from each other the evanescent fields of the guided modes overlap, resulting in the formation of a symmetric mode with a larger $n_{\text{eff}}$ and an anti-symmetric mode with a smaller $n_{\text{eff}}$. By adjusting the inter-membrane separation the effective index can be controlled and tuning of the cavity resonances is expected.\cite{1}
To estimate the amount of shift as a function of the distance, a three-dimensional finite element method (FEM) simulation has been performed to calculate the spectrum of a cavity on two In$_{(1-x)}$Ga$_{x}$As$_{y}$P$_{(1-y)}$ ($x=0.26$, $y=0.57$, $n=3.42$) slabs separated by an air layer. A thickness of 180 nm for both membranes has been chosen to have only two guided modes in the double waveguide.
To provide an example of tuning, an L3 cavity (three holes missing) on a triangular lattice, having a fill factor FF$=35\%$ and a lattice constant $a=420$ nm, is considered. Fig.~\ref{fig:theory}(b) shows the calculated wavelengths of the L3 y-polarized symmetric and anti-symmetric modes as a function of the separation $d$. A maximum tuning of 200 nm can be achieved for both modes when the entire range is considered. The calculated tuning rate $\delta\lambda/\delta d$ approaches the value of 1 nm/nm for $d < 150$ nm showing strong optomechanical coupling\cite{12} and high displacement sensitivity.\cite{18}
A well-known limit for electrostatic actuators is the pull-in effect which, for a simple plate attached to a spring, limits the maximum stroke to 2/3 of the distance between the electrodes at rest. Therefore it is desirable to work at small initial gaps ($d\sim100$ nm), corresponding to higher tuning rates, to maximize the total mode shift before pull-in. Such a small distance poses extreme fabrication challenges for the release of the structures, easily leading to stiction failures and bending due to residual strain. To successfully fabricate a device, a conservative value of $d_0 = 240$ nm, corresponding to a maximum theoretical tuning range of 20 nm, has been used.

\begin{figure}
\includegraphics{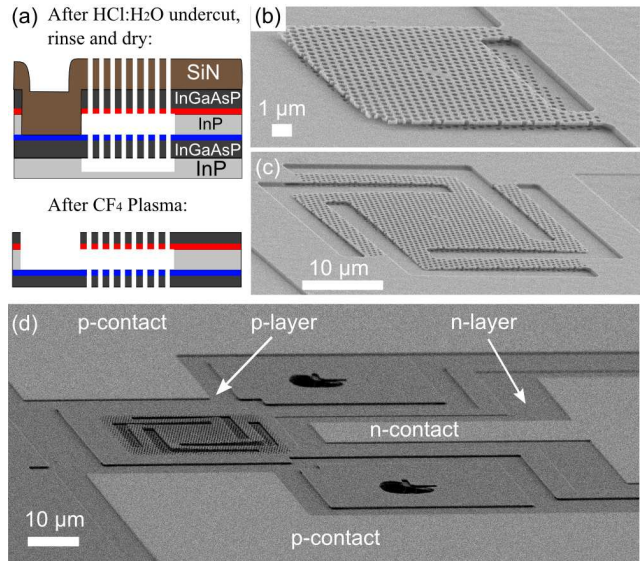}%
\caption{\label{fig:sem} (a) Process used to realize photonic crystals in freestanding membranes. The p- and the n-layers are highlighted in blue and red, respectively. (b) SEM micrograph of a 14$\times$14 $\mu$m cantilever realized with this technique. Other shapes have been designed like (c) a 4-arms suspended membrane. (d) Side view of the complete device showing electrodes, doped layers and the photonic crystal NEMS.}%
\end{figure}

Samples are grown on an InP substrate by metal organic chemical vapor deposition and consist of a 50 nm InGaAs etch stop layer, 1 $\mu$m InP buffer layer and two, 180 nm thick, In$_{(1-x)}$Ga$_{x}$As$_{y}$P$_{(1-y)}$ ($x=0.26$, $y=0.57$) layers separated by a 240 nm thick InP. The epitaxial structure is terminated by a 50 nm InP capping layer. At the center of the upper membrane a single layer of self-assembled InAs QDs emitting at 1550nm at 300K is grown.\cite{11} The bottom (top) 50 nm thick layer of the upper (lower) InGaAsP membrane is n-(p-)doped to $10^{18}$ cm$^{-3}$. In this configuration the QDs, placed in an undoped region above the junction, should not be affected by the applied voltage. 
As a first step of the process, the InP capping layer is chemically etched. Vias are opened to the p- and n- doped layers using $\text{CH}_4$:$\text{H}_2$ reactive ion etching and standard lithographic techniques. In the etching of the vias to the p-membrane mechanical structures with various shapes are defined as well. Ti/Au (50/200 nm) contacts are evaporated on the doped layers and lifted-off in acetone. 
To realize the photonic crystal holes a thick (400 nm) $\text{Si}_3\text{N}_4$ mask is used. The pattern is defined by Electron Beam Lithography and transferred in the nitride. The holes are then deeply etched by inductively coupled plasma (ICP) using $\text{Cl}_2$:$\text{Ar}$:$\text{H}_2$ chemistry at 200$^\circ$C. The InP sacrificial layer between and under the membranes is removed by a solution of hydrochloric acid in water at 2$^\circ$C. A major fabrication issue is the collapse of the upper membrane due to strong capillary forces (stiction) in the rinsing liquids (de-ionized water and boiling isopropanol) during drying. In NEMS fabrication, stiction is usually avoided by CO$_2$ supercritical drying.\cite{23} Here we introduce a different method which does not require the high-pressure chambers needed for supercritical drying. Indeed we use the SiN etching mask as a mechanical support during the drying (performed in isopropranol vapor and air) and then remove it using a CF$_4$ plasma in a barrel etcher. Since no liquid is involved in the last step, no capillary forces exist and stiction rarely occurs. Moreover no damage on the metal contacts or on the InGaAsP surface due to $\text{CF}_4$ plasma has been observed. Fig.~\ref{fig:sem}(b and c) show scanning electron microscope (SEM) images of the free-standing NEMS after the silicon nitride mask has been removed. Residual strain is observed causing the structures to bend slightly towards the substrate. This effect might limit the actual travel range of the actuator.
The final device with vias and contacts is shown in Fig.~\ref{fig:sem}(d). 

\begin{figure}
\includegraphics{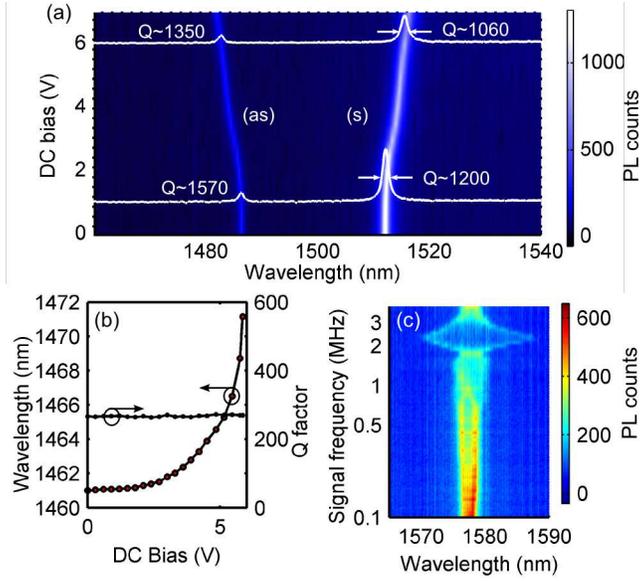}%
\caption{\label{fig:results} (a) PL of symmetric (s) and anti-symmetric (as) modes of an L3 cavity obtained sweeping the DC bias. (b) Peak shift of an H1 dipole mode. No change in the Q factor is observed (c) Spectrum of an L3 mode obtained by applying a 6 V peak-to-peak sinusoidal signal with a 4 V DC offset at various frequencies.}%
\end{figure}

A room temperature micro-photoluminescence ($\mu$PL) setup is used to characterize the fabricated devices. A 785 nm continuous wave diode laser is focused by a microscope objective (NA 0.4) into the sample to excite the QDs and use them as an internal light source. The PL signal is collected through the same objective, coupled into a fiber and analyzed using a spectrometer. The setup is equipped with micro probes mounted on a movable stage and connected to a signal generator. The measurements are performed by acquiring PL spectra (integration time $\sim 10$ s) from each device without bias and then gradually increasing the input voltage. 
Figure \ref{fig:results}a shows the PL spectra of an L3 cavity ($a=500$ nm, FF$=35\%$) on a spiral-shaped cantilever (Fig.~\ref{fig:sem}c) as a function of the applied bias. Two modes are visible when no voltage is applied, corresponding to the symmetric (s) and anti-symmetric (as) y-polarized modes of the cavity. The deviation between the measured wavelengths and theory is likely due to fabrication imperfections. When voltage increases the as-mode blue-shifts and the s-mode red-shifts as expected. The maximum tuning is 4.8 nm and 4.4 nm at 6.5 V, respectively. At higher bias the diode reaches breakdown region, the current rapidly increases and no more tuning is observed. The Q factor of both modes is not significantly degraded during tuning. The small broadening of the peaks is probably due to fluctuations of the actuator, causing small shifts of the mode during the spectrum acquisition. The sweep can be repeated several times and only small hysteresis (about 5\% of the original tuning) during the first 3 cycles has been measured, after which the tuning is reproducible. The fact that a simultaneous blue and red shifting can be seen is a clear proof that tuning is not due to Joule heating, which could be produced by the currents in the device, but to the mechanical displacement of the membranes.
In this device, the electrostatic actuation is limited by the voltage drop in the contact layer. Indeed, the current-voltage characteristic of the junction shows relatively high current in reverse bias (10 $\mu$A at 6 V) which can be partly explained by the diffusion of p-dopants through the intrinsic InP layer and partly to the large area of the junction itself. Sheet resistances up to 10 k$\Omega/$square have been measured for the 50 nm p-layer, causing voltage drops and reducing the effective electric field across the air gap. Higher forces may be obtained by improving the design and the fabrication. 

In devices where the reverse current is lower we have observed a larger tuning. For example, Fig.~\ref{fig:results}b shows 10 nm tuning of the dipole mode (symmetric, red-shifting) of an H1 (one hole missing, $a=530$ nm, FF$=35\%$) cavity when DC bias is increased in small steps up to 5.8 V. At higher voltages instability due to pull-in occurs and the cavity mode disappears from the spectrum. This has been confirmed by analyzing the device with a SEM after the membranes have collapsed. The theoretical value of 20 nm has not been observed probably because of the bending of the suspension arms (Fig.\ref{fig:sem}c) which might cause pull-in before the central part of the structure reaches 2/3 of the separation at rest. 

To provide a further evidence of mechanical tuning, the device has been operated with a sinusoidal signal. Using a DC offset of 4 V and an AC drive signal with an amplitude of 6 V, a PL spectrum is acquired as a function of the signal frequency.  Fig.~\ref{fig:results}c shows the results of such an experiment for an L3 cavity mode on a cantilever. When a mechanical resonance (at 2.4 MHz in the graph) is hit, the displacement is amplified resulting in a broadening of the cavity peak. The maximum amplitude of the tuning is as high as 15 nm. From the spectra it is also possible to estimate a mechanical quality factor $Q_m \approx 2$ for this resonance. Since experiments are performed at ambient pressure a high damping is expected. The resonance is not due to the fundamental vibrational mode of the tested structure, which is expected to be around 350 kHz. We think that such a mode is highly damped and therefore not visible, while the observed broadening is caused by the second longitudinal mode, theoretically predicted around 2.5 MHz. 

In conclusion, the mechanical displacement of double membrane NEMS has been confirmed by the observation of the anti-symmetric mode blue-shift ($\sim 4$ nm), the collapse due to pull-in and the resonance when operated with an AC signal. The Q factors of the cavities are currently limited by simple designs of the cavities and not intrinsic to the double membrane structure.\cite{1} The device can be still improved by reducing the reverse bias currents in order to reach pull-in before diode breakdown, by reducing the inter-membrane distance to increase the tuning range and by realizing advanced designs for pull-in free operation. 

We acknowledge useful discussions with H. Ambrosius, L. Balet, R. van der Heijden and F. Pagliano. This research is supported by the Dutch Technology Foundation STW, applied science division of NWO, the Technology Program of the Ministry of Economic Affairs and the Dutch HTMS program.

\newpage

\end{document}